\documentclass[11pt]{article}
\usepackage{amssymb,latexsym,amsmath}
\usepackage{latexsym,amssymb,amsfonts,epsfig,graphicx,color}

\newcommand{\be}{\begin{equation}}
\newcommand{\ee}{\end{equation}}
\newcommand{\bea}{\begin{eqnarray}}              
\newcommand{\eea}{\end{eqnarray}}

\title{Gauge-free Coleman-Weinberg Potential}
\author{Srijit Bhattacharjee$^a$, Parthasarathi Majumdar$^b$\\
$^a$Astroparticle Physics \& Cosmology Division\\Saha
Institute of Nuclear Physics\\Kolkata 700 064, India.\\
$^b$Department of Physics,\\ Ramakrishna Mission Vivekananada University\\
Belur Math, Howrah 711202, India.}
\date{}
\begin{document}
\maketitle

\begin{abstract}

The gauge-dependence of the one loop Coleman-Weinberg effective
potential in scalar electrodynamics is  
resolved using a gauge-free approach not requiring any gauge-fixing of
quantum fluctuations of the photon degrees of freedom. This
leads to a unique dynamical ratio at one loop of the Higgs mass
to the photon mass. We compare our approach and results with
those obtained in geometric framework of DeWitt and Vilkovisky, which
maintains invariance under field redefinitions as well as invariance
under background gauge transformations, but {\it requires}, in
contrast to our approach, gauge fixing of {\it fluctuating} photon fields. 
We also discuss possible modifications of the Coleman-Weinberg potential
if we adapt the DeWitt-Vilkovisky method to our gauge-free approach
for scalar QED.  
\end{abstract}


\section{Introduction}

The Coleman-Weinberg effective potential has been a very important
tool to study the vacuum structure and radiative Higgs mechanism for
mass generation in gauge field theory. Since the
seminal paper by Coleman and Weinberg \cite{cw}, a lot of effort have
been made to calculate the effective potential in a systematic
manner. One of the most significant works in this context was made by Jackiw
\cite {jacki} where the one loop effective action is computed using loop-expansion
in a purely functional integral
scheme. It is also shown in this work that the one loop effective potential is gauge
{\it dependent}, and in fact could be gauged away within a class
of gauges by appropriate choice of a gauge parameter . The issue of gauge
dependence of the effective action has been extensively analyzed in the past [see \cite{Huggins} and the
references therein]. It has also been pointed out that the 
effective action is not only gauge dependent but also depends upon 
field reparameterization  \cite{Vilk}. A new 
approach for computing the effective action was introduced by DeWitt and Vilkovisky
\cite{Vilk,DeWitt}, where the effective action was explicitly shown to
be reparameterization  independent. Further, this DeWitt-Vilkovisky
scheme used a background field method and the effective action is
explicitly background gauge invariant, even though it requires a choice
of gauge to integrate the fluctuating part of the
gauge fields. 

Our aim in this paper
is to recalculate the effective potential for scalar QED at one loop,
using an approach which we call {\it gauge-free}. In this framework, 
quantum fluctuating dynamical variables are
{\it manifestly inert}  under (abelian) gauge transformations \cite{gfpm},
\cite {bm2}. In contrast to the usual treatment of functional quantization of gauge
theories involving the Faddeev-Popov ansatz, here we propose a reformulation of
electrodynamics in terms of a {\it physical} vector potential entirely
free of gauge ambiguities right from the outset \cite {gfpm}, {\bf and which is spacetime
divergenceless} : $\partial
\cdot {\bf A}_{\cal P} =0$. It is important to note that this last property of the vector
potential is {\it not} the Lorentz gauge condition but a {\it physical restriction} on
a physical vector potential. It is merely a restatement of the fact that in the standard formulation of pure electrodynamics, gauge transformations act only on the unphysical
degrees of freedom of the gauge potential, with the physical, gauge invariant part of
the gauge potential being divergenceless by definition, not as a
matter of choice. The gauge free approach, by virtue of being
based on a physical, divergenceless vector potential, evades the
entire issue of gauge redundancy. Quantizing the theory with this prescription leads to a propagator that is gauge invariant by construction, in contrast to the standard
photon propagator.

The charged matter fields can be coupled with this
{\it physical} photon field in a gauge-free fashion if we rewrite the
fields in polar representation, so as to `separate' charge and spin
degrees of freedom. The modulus of the matter fields
carries  the spin (scalar) degrees of freedom and the phase part carries only
the charge of it. This separation of spin and charge actually enables
us to represent the theory in terms of manifestly gauge-inert
variables. 

We would like to mention here that
the radial decomposition of charged matter fields sometimes been
referred to as imposing `unitarity gauge', in the literature \cite{Gross_Jackiw,Dolan_Jackiw}. We
do not agree with this notion because ``unitarity gauge'' is not
really a
choice of gauge in the sense of other gauge choices, but a unique representation of a gauge theory with
redefined fields which do not transform under gauge
transformations. In our gauge-free approach for the case of scalar
QED, the action has an apparent similarity 
to the one obtained by Dolan-Jackiw by employing the so-called
unitarity gauge. However, there is a  crucial difference between their approach and
ours in the functional integral. We include the physical constraint on the vector potential $\partial
\cdot {\bf A}_{\cal P} =0$ in the functional integration which is absent in \cite{Dolan_Jackiw} (See section 3). In the same
spirit, we have also shown earlier that the Higgs-weak vector boson
sector of the standard electroweak theory can also
be rewritten in terms of manifestly gauge inert degrees of freedom
\cite{bm2}.  The functional quantization of scalar quantum
electrodynamics leads at the quantum level to a one loop effective
potential which realizes the Coleman-Weinberg mechanism of mass
generation in a gauge-free framework, thus resolving the issue of  its
gauge dependence. However, since  the reparameterization invariance
can only be ensured by treating the theory in the DeWitt-Vilkovisky (DV)
approach, we have calculated the gauge-free theory according  to the
DV approach and get a different result from the one calculated earlier
by Kunstatter \cite{Kun}. This difference is an indication that by
eschewing redundant field degrees of freedom from the outset, it is
possible to obtain a unique result for the vector potential.  

We may mention that there have been many efforts in the past towards identifying 
gauge invariant variables and formulating gauge theories in terms of
these. See e.g. the recent paper by Ilderton et. al. \cite{ild} which provides
a definitive guide to the literature of the mid-1990s on these efforts, including the
authoritative contribution of Lavelle and McMullan \cite{Mcmullan}. Related to this earlier work,
recently Niemi et. al. \cite{nie} and Faddeev \cite{fadd} have proposed a
gauge invariant description of the Higgs-gauge
sector of standard electroweak theory whereby the Higgs field is given a novel
interpretation as the dilaton in a {\it conformal} curved background. However, in a completely gauge-free framework
it has been shown that the novel interpretation of Niemi et. al. doesn't survive under quantization. Genuine one-loop
effect actually cancels out the contribution of the dilaton field \cite{bm2}. 

Although similar in spirit to some of these assays in a broad sense,
we reiterate that our approach is
distinct in that it is formulated in terms of a {\it local,
physical} vector potential (instead of field strengths) as a {\it
fundamental field variable}. In other words, we
propose an alternative action/field equations as a new starting point rather than attempt to
express the standard gauge theory action in terms of new variables. 

The paper has been organised as follows: In the next section a brief
review on different approaches to calculating the effective potential
is given and the relative advantages of the gauge-free
approach is discussed. In Section 3 we motivate our gauge-free
approach by describing the functional quantization of vacuum
electrodynamics. Then in Section 4, we deal
with charged scalar fields (as already mentioned) and study the abelian
Higgs mechanism. In Section 5 we consider the Coleman-Weinberg perturbative mass generation
mechanism and obtain a gauge ambiguity free mass spectrum. Section 6 is
devoted to a brief description of the DV method and a
unique one loop effective-potential is derived adopting a combination of
these two approaches.  In Section 7 we generalize our gauge free approach to
symmetric and  anti-symmetric second rank tensor fields. We conclude in section
8 with a brief discussion on Yang Mills fields without Higgs scalars.   

\section{Effective Potential from different approaches}

The dependence of effective potential on choice of gauge was first shown by Jackiw \cite {jacki} in the context 
of scalar-QED. The one loop effective potential for scalar QED obtained by Jackiw in a general gauge
 $-\frac{1}{2\alpha}(\partial_{\mu}A^{\mu})^2$ is gauge dependent:
 
\begin{equation}
 V_{eff}(\phi_c)=\frac{\phi_c^4}{4!}\left[\lambda_R+\frac{\hbar}{8\pi^2}\left(\frac{5}{6}\lambda_R^2+9e^4-\alpha e^2\lambda_R\right)
\ln{\phi^2_c}\right] \label{jac}
\end{equation}
It is thus possible to gauge away the one loop contribution to the
effective potential by choosing $\alpha$:  \[\alpha=\frac{5\lambda_R}{6e^2}+\frac{9e^2}{\lambda_R}\]
This is a serious problem because this raises the question of physicality  of 
the effective potential itself. Soon after this
result, Dolan and Jackiw \cite{Dolan_Jackiw} calculated the same
effective potential in the so-called unitarity gauge and asserted that the 
theory in this gauge has no unphysical degrees of freedom, and hence
the effective potential is {\it physical}. It is given by 
\begin{eqnarray}
 V_{U}&=&{1\over 2}dm^2\left(1-\frac{\hbar\lambda}{64\pi^2}\right)\rho_c^2~+~{\lambda \over 4!} \rho_c^\\ &+&
\frac{\hbar\lambda}{64\pi^2}\left[3e^4\rho_c^4\ln(\frac{\rho_c^2}{m^2})~+~(m^2+\lambda/2\rho_c^2)^2\left(\ln(1+ \frac{\lambda\rho_c^2}{2m^2}
\right)\right] , \label{uni}
\end{eqnarray}
and is clearly different from (\ref{jac}). 

The problem of a non-unique one-loop effective potential had been discussed in several papers \cite{Kun},
\cite{kun-leivo}. It has been shown that the one loop effective potential depends not only upon the choice of gauge but also on the 
reparameterization  of the fields. The reparameterization  invariance means that the effective action coincides with the original
one under a field redefinition  $\phi \rightarrow
\phi^{'}$. DeWitt-Vilkovisky \cite{Vilk,Toms-book} have introduced an effective 
action formalism which addresses both the problems (dependence of effective action on gauge fixing condition and on reparameterization of the fields) and provides a reasonable solution ! A lot of work have been done on the issue of gauge and parameterization 
dependence of the effective potential and notable amongst those is the
result obtained by Kunstatter \cite{Kun} in DV approach.
 \begin{eqnarray}
V_{eff}(\rho_c) &=& {\lambda \over 4!} \rho_c^4 + {\hbar \over 64\pi^2} \left(3e^4 + \frac{5}{18} \lambda^2+ \frac23 
\lambda e^2 \right)\rho_c^4\left[\log {\rho_c^2 \over M^2} - {25\over 6} \right]~.
\label{rveff}
\end{eqnarray}

However, DeWitt-Vilkovisky's method does indeed need gauge fixing of the fluctuating gauge degrees of freedom which are being integrated over in the partition
function. This of course is easily obviated by the use of the gauge
free approach adopted in the present paper. Thus, the calculation of
the effective potential becomes much easier in the gauge-free DV
approach since the unphysical electrodynamic degrees of freedom are absent
from the outset, and all fields are manifestly inert under $U(1)$ gauge transformations. 
Later we shall explicitly calculate the one-loop effective potential
for scalar QED in the gauge-free DV approach and show that the 
potential does get modified from our earlier result and it also
differs from the one obtained by Kunstatter \cite{Kun}. 

\section{Gauge Free Vacuum electrodynamics}

We start the with the Maxwell Electrodynamics to develop the idea of gauge-free quantization. For the standard gauge potential (abelian) one form $A$ and semi-infinite curve $C$ from spatial infinity to $x$,
\bea {\bf A}_C (x) &\equiv& h_{C(\infty,x)} [A] (A + d) (h_{C(\infty,x)} [A])^{-1}\\
&=& A - d\int_{C(\infty,x)} A 
\eea

For another semi-infinite oriented curve $C^{'}$ from $\infty$ to the point, it is easy to see that

\bea
{\bf A}_C - {\bf A}_{C^{'}}&=& d\left[\int_{C^{'}}-\int_{C}\right] A\\
&=&d\int_{S_{CC^{'}}} dA
\eea
where, the second line follows from the first by Stoke’s theorem with $S_{CC^{'}}$ being the surface
bounded by the two semi-infinite curves C and C from spatial infinity to x. The right hand side of the
second equation vanishes because $d^2 = 0$. In other words, even though $A_C$ formally depends
on the curve C and is expected to be non-local, actually it is independent of C and hence local.
This also agrees with the fact the $A_C$ is gauge-independent $\forall$ C. Hence we drop the subscript C in what follows.
Now, observe that

\be {\bf A}= A - d\int_C A=A-d\int d^4x^{'} dy_a A^a(x^{'})\delta^4(y-x^{'})
\ee

Define the d'Alembert Green function $G(x, x^{'} )$ as $\Box G(x, x^{'} ) = \delta^4 (x- x^{'} )$. This is consistent
with 
\be \partial_a G(x, x^{'})=\int_{C(\infty,x)} dy_a \delta^4 (x - y)\ee

as can be seen by taking partial derivatives on both sides. Substituting eq (6) in (5), and
performing a partial integration, we get
\be
{\bf A}=A - d\int d^4 x^{'} G(x, x^{'} ) \partial^{'} · A(x^{'}) ={\bf A}_{\cal P}
\ee

Here ${\bf A}_{\cal P}$ denotes a spacetime transverse physical vector which can easily be proved as follows.

\bea{\bf A}_{\cal P}&=&A - d\int
d^4 x^{'} G(x, x^{'} ) \partial^{'} · A(x^{'})\nonumber\\ \nonumber 
\partial.{\bf A}_{\cal P}&=& \partial.A\,-\,\Box\int d^4x^{'}\partial^{'}.A(x^{'}) G(x-x^{'})\\
&=&0
\eea


With this physical field ${\bf A}_{\cal P}$ we now write an action for Maxwell Electrodynamics.

\begin{equation}
 \partial \cdot {\bf A}_{\cal P} =0 \label{trans}
\end{equation}
We emphasize the fact that no gauge fixing needs to be employed here; but since the functional integral
describing the vacuum-to-vacuum amplitude is over all configurations
of the vector field ${\bf A}_{\cal P}$, the transversality constraint
must be directly inserted into the integral to ensure
that the integral is only over transverse field configurations. The gauge-free formulation for vacuum electrodynamics 
starts with the action 
\begin{eqnarray}
S[{\bf A}_{\cal P},\Lambda;{\tilde {\bf J}}] &=& \int \left[ -\frac12~\partial_{\mu}
  A_{{\cal P}\nu}~\partial^{\mu}A^{\nu}_{\cal P} + {\tilde J} \cdot A_{\cal P} + \Lambda \partial
  \cdot A_{\cal P} \right] ~ \nonumber \\
&=& \int \left[-\frac12~\partial_{\mu}  A_{{\cal
      P}\nu}~\partial^{\mu}A^{\nu}_{\cal P} + J \cdot A_{\cal P} \right] ~. \label{actn}
\end{eqnarray}

The second line above follows from the first by eliminating the Lagrange
multiplier field $\Lambda$ through its equation of motion and defining
$J_{\mu}$ such that $\partial \cdot J =0$. The relevant
vacuum-to-vacuum amplitude (in presence of a transverse source) is given by
\begin{eqnarray}
Z[{\bf J}] ~&=&~ \int {\cal D}{\bf A}_{\cal P}~ \exp i \left(~ \frac12
~A_{\cal P}^{\mu}~ \Box~A_{{\cal P} \mu}   ~+~ \int d^4x~{\bf J}
\cdot {\bf A}_{\cal P} \right) ~ \delta[\partial_{\mu}
~{\bf A}_{\cal P}^{\mu}] \nonumber \\
& ~=~& \int {\cal D}{\cal S}~{\cal D}{\bf A}_{\cal P}~\exp i~\int d^4x~ \left [ \frac12
~A_{\cal P}^{\mu}~ \Box~A_{{\cal P} \mu} ~+~ ( J_{\mu} ~-~ \partial_{\mu} ~{\cal
  S}) ~A_{\cal P}^{\mu} \right] ~. ~\label{vacam}
\end{eqnarray}

In the second line of (\ref{vacam}) we have introduced an auxiliary
scalar field ${\cal S}$ which acts as the Lagrange multiplier for the
physical constraint (\ref{trans}). After integrating over ${\bf A}_{\cal P}$ and auxiliary field we get,
\begin{eqnarray}
Z[{\bf J}] &~=~&N\exp{\frac{-i}{2}\left[J^{\mu}\left(\frac{\eta_{\mu\nu}}{k^2}-\frac{k_{\mu}k_{\nu}}{k^4}\right)J^{\nu}\right]}  \label{integr}
\end{eqnarray}
A series of
partial integrations and using the transversality of the current density ${\bf
J}$, and also identities like $\partial_x {\cal G}(x-y) = - \partial_y {\cal
D}(x-y)$ leads to this simple expression

It is now straightforward to extract the free photon propagator from
eq. (\ref{integr}):
\begin{eqnarray}
{\cal D}_{\mu \nu}(k) &~\equiv~& \frac12 ~{\delta^2 W[{\bf J}] \over \delta
 J^{\mu}(k) ~\delta J^{\nu}(-k) }\Bigg|_{{\bf J}=0}\\
&~=~&\frac{1}{k^2 +i\epsilon}\left(\eta_{\mu\nu}-\frac{k_{\mu}k_{\nu}}{k^2~+~i\epsilon}\right)
\end{eqnarray}
Clearly, this propagator does not possess any gauge
artifacts. Where we have introduced the generating functional for connected Green's function via $W[{\bf J}]=-iLog{Z[{\bf J}]}$

 In our gauge-free formulation, spacetime
divergencelessness is {\it not} a matter of choice, it is a defining
feature of what we mean by electromagnetism. Finally, note also that the free
photon propagator falls off as $1/k^2$ for large momentum, as is expected for
a {\it local} field.  

\section{Gauge-free electrodynamics with sources}

\subsection{Charged matter fields}

All charged matter fields are complex fields ${\Phi}$
such that they can be `radially' decomposed : ${\Phi} = {\phi} \exp i\theta$ where 
${\phi}$ carries all the spin degrees of freedom of ${\Phi}$ and the
phase field $\theta$ is a scalar field which appears in the action only
through its first order derivative $\partial \theta : S[\Phi] =
S[\phi, \partial\theta]$. The gauge-free prescription for coupling the
gauge-free vector potential ${\bf A}_{\cal P}$ to $\Phi$ is exceedingly simple
: leaving $\phi$ as it is in the action, simply replace $\partial \theta
\rightarrow \partial \theta - e A_{\cal P}$, so that $S[\Phi] \rightarrow
S[\phi, \partial \theta - e A_{\cal P}] + S_{free} [A_{\cal P}]$. Recall of
course that the gauge-free ${\bf A}_{\cal P}$ is subject to the
4-divergencelessness constraint (\ref{trans}). The interaction with matter for
this vector potential is merely to add a {\it physical longitudinal} part to
it so that potentially it can now turn massive even in the weak coupling
limit, depending upon the form of $S[\Phi]$. An example of this is the Abelian
Higgs model of scalar electrodynamics \cite{aspc}.  

\subsection{Abelian Higgs Model}

A charged scalar admits the radial decomposition $\phi = (\rho/\sqrt{2}) \exp i\theta$
where $\rho$ and $\theta$ are both to be treated as physical fields. With this
decomposition, the action of the complex scalar field appears as (suppressing
obvious indices) 
\begin{eqnarray}
S_0[\rho, \theta] = \int d^4 x \left[\frac12 (\partial \rho)^2 + \frac12 \rho^2 (\partial
  \theta)^2 - V(\rho) \right] ~. \label{comsc}
\end{eqnarray}
This action (\ref{comsc}) is invariant under the global $U(1)$ transformations $ \rho
\rightarrow \rho~,~ \theta \rightarrow \theta + \omega$ where $\omega$ is a
real constant. 

Observe now that one can define $\Theta \equiv \theta - e a$ where $a$
is introduced as part of the standard $U(1)$ gauge potential which
carries the entire gauge transformation $a \rightarrow a + e^{-1}
\omega$ when one couples the scalar theory to the standard gauge field
$A_{\mu}$. It is obvious that $\Theta$ is invariant under gauge
transformations. Following our prescription above, coupling to the physical electromagnetic
vector potential is obtained through the action (dropping obvious indices)
\begin{equation}\label{abh2}
S[\rho,\Theta,{\bf A}_{\cal P}] = \int d^4x \left[ \frac12 (\partial
\rho)^2 + \frac12 e^2 \rho^2 ({\bf A}_{\cal P} - e^{-1}\partial \Theta)^2
~- \frac12 (\partial A_{\cal P})^2 - V(\rho) \right] 
\end{equation}
where $V(\rho)$ is the scalar potential, and ${\bf A}_{\cal P}$
obeys the divergenceless constraint (\ref{trans}). 
It is interesting that the phase field $\Theta$
occurs in the action only through the combination ${\bf A}_{\cal P} - e^{-1}
d\Theta$; this implies that the shift $\Theta \rightarrow \Theta +
const.$ is still a symmetry of the action. However, since there is no
canonical kinetic energy term for
$\Theta$, it is hard to associate a propagating degree of freedom with
$\Theta$. Indeed, if one first makes a field redefinition
\begin{eqnarray}
Y_{\mu} ~\equiv~ A_{{\cal P} \mu}~ -~e^{-1}~ \partial_{\mu}~\Theta ~. \label{ugau}
\end{eqnarray}
the $\Theta$ can be completely absorbed into the new vector field ${\bf
  Y}_{\mu}$, appearing only in the constraint which replaces (\ref{trans}) 
\begin{eqnarray}
\partial \cdot Y = - \Box \Theta ~. \label{try}
\end{eqnarray}

This implies that ${\bf Y}$ has three {\it physical} polarizations rather than the two that
${\bf A}_{\cal P}$ had. However, this does not immediately imply that ${\bf
Y}$ has acquired a mass. Upon eliminating $\Theta$ through the constraint
(\ref{try}), eq. (\ref{abh2}) assumes the form
\begin{equation}\label{abhm}
S[\rho,{\bf Y}] = \int d^4x \left[ \frac12 (\partial
\rho)^2 + \frac12 Y^a \left( (\Box + e^2\rho^2) \eta_{ab}
  -\partial_a \partial_b \right) Y^b - V(\rho) \right], 
\end{equation}
This is the gauge-free Abelian Higgs model. 

One can now think of two kinds of scalar potentials $V(\rho)$: one for
which the minimum of the potential $\langle \rho \rangle =0$ and the
other for which the minimum lies away from the origin $\langle \rho
\rangle = \rho_c \neq 0$. It is this second case which is of interest to us.
If $V(\rho)$ has a minimum at $\rho = \rho_c \neq 0$
one now also defines $\rho \rightarrow \rho +
\rho_c$, it is easy to see that the ${\bf Y}$ acquires a mass
$m_{\bf Y}^2 = e^2 \rho_c^2$ while the $\rho$ also acquires a mass
$m_{\rho}^2 = V''(\rho_c)$. This is precisely the manner in which a {\it
physical} longitudinal degree of freedom conjoins the photon field to produce
a massive vector boson. In doing so, the new vector
potential ${\bf Y}$ is no longer subject to the transversality constraint (\ref{trans}). It
thus has one degree of freedom more than the ${\bf A}_{\cal P}$. Observe that
the Higgs phenomenon of mass generation {\bf did not involve any symmetry
breaking at all}, reminding us of Elitzur's theorem \cite{elit} proved for QED on a
cubic lattice. The vacuum expectation value $\rho_c \equiv \langle \rho \rangle$ does not break any
continuous symmetry at all. {\it The Higgs mechanism is a gauge-free mechanism
of mass generation, involving neither symmetry breaking of any sort, nor
unphysical particles in the spectrum}. 

Before closing this subsection, we point out that this aspect of the phase
field attaching itself to the photon field as a {\it physical}
longitudinal piece, is not confined to charged scalar fields. Consider for
instance a free charged Dirac field given by the action 
\begin{eqnarray}
S[\psi] = \int d^4 x~ {\bar \psi} (i \gamma \cdot \partial -m) \psi
~.\label{frd1}
\end{eqnarray}
Performing the `radial decomposition' $\psi = \chi \exp i\theta$ this reduces
to 
\begin{eqnarray}
S[\chi, \theta] = \int d^4 x \left( ~{\bar \chi}(i \gamma \cdot \partial -m)
  \chi - {\bar \chi} \gamma \cdot \partial \theta \chi~ \right) ~. \label{frd2}
\end{eqnarray} 
This action is of course invariant under the global $U(1)$ transformations
$\chi \rightarrow \chi~,~ \theta \rightarrow \theta + \omega$ for a constant
$\omega$. Employing our prescription above for coupling this field to the physical
electromagnetic vector potential, we notice that the action now reads
\begin{eqnarray}
S[\chi, \theta] = \int d^4 x \left( ~{\bar \chi}(i \gamma \cdot \partial -m)
  \chi - {\bar \chi} \gamma 
\chi \cdot (\partial \theta - e A_{\cal P})~ \right) ~. \label{frd3}
\end{eqnarray}
It is obvious from the above that under any interaction, the vector potential
is {\it poised} to pick up a physical longitudinal piece ($\partial \theta$)
corresponding to the `charge mode'. However, in this case there is no
mechanism (at tree level) of mass generation due to the absence of a `seagull'
term. But this could be an artifact of weak coupling. In the 1+1 dimensional
quantum electrodynamics model analyzed half a century ago by Schwinger \cite{schwi}, the
photon field does pick up a manifestly gauge invariant mass as an exact dynamical result.

\section{Gauge-free scalar QED: Coleman-Weinberg Mechanism}

The Coleman-Weinberg mechanism \cite{cw} is a radiative mechanism whereby a
scalar electrodynamics theory with massless photons and charged scalar bosons, changes
its spectrum due to perturbative quantum corrections. Both the neutral
component of the scalar boson and the vector boson acquire physical masses given by the
parameters of the theory. In its incipient formulation, the mechanism has been
shown to be gauge-dependent \cite{jacki}, thereby casting doubt on its
physicality. Using the gauge free reformulation given above, we compute in
this section the one loop effective potential of the theory, and argue that
the effect is physical at this level.

The action for the theory is already given above eq.(\ref{abh2}), with the
choice $V(\rho) = (\lambda/4!) \rho^4$. Following
\cite{cw}, the theory is
quantized using the functional integral formalism. In the standard formulation
of QED, one needs to resort to the Faddeev-Popov technique of gauge fixing and
extracting the infinite volume factor associated with the group of gauge
transformations, from the vacuum persistence amplitude (generating functional
for all Green's functions), in
order that this amplitude does not diverge upon integrating over gauge
equivalent copies of the gauge potential. In the gauge free approach here, this
technique is not necessary. The integration over the transverse gauge
potential is, of course, restricted to configurations that obey the spacetime
transversality condition (\ref{trans}). Since the integration variables are
unambiguous, the task, at least at the one loop level, is simpler.

The generating functional is thus given by
\begin{eqnarray}
Z[J, J', {\bf J}] &=&e^{{i \over \hbar} W[J, J', {\bf J}]}= \int {\cal D}\rho ~{\cal D}\Theta ~{\cal D}{\bf A}_{\cal
  P}  ~\exp {i \over \hbar} \left[~S[ \rho,\Theta,{\bf A}_{\cal P} ]
+ \int d^4x(J \rho + J' \Theta + {\bf J} \cdot {\bf A}_{\cal P})~ \right]
\nonumber \\
&&~ \cdot \delta[~\partial_{\mu}~A^{\mu}_{\cal P}~] ~. \label{vpam}
\end{eqnarray}
Here, the integration measures ${\cal D} \rho = \Pi_x d \rho(x)~,~{\cal D}{\bf
A}_{\cal P} = \Pi d{\bf A}_{\cal P}$, but the remaining measure ${\cal
D}\Theta = Det \rho^2 \Pi_x d\Theta(x)$. The extra factor of $Det \rho^2$
can be seen to arise if one begins with the generating functional first
expressed as functional integrals over a complex scalar field and its complex
conjugate. Alternatively, one can obtain the configuration space functional
integral starting with the functional integral over phase space. Integration
over the momentum conjugate to $\Theta$ produces the same factor \cite{senj}.

Indeed, it is a similar factor which has been interpreted in \cite{fadd} as
representative of a background spacetime which is {\it conformally} flat,
rather than flat, with the `radial' component of the Higgs field $\rho$ playing the
role of the conformal mode. In \cite{nie}, a slightly different interpretation is
given of this radial Higgs field as a {\it dilaton} field. Formally, there is
indeed novelty in both interpretations. However, when perturbative effects are
included, at least at the one loop level, such interpretations will be seen to
be in need of modification to account for scaling violations due to renormalization \cite{bm2}.  

The effective action $\Gamma[\Phi]$ which is the generating functional for one particle
irreducible diagrams (1PI), is generically defined as usual through the Legendre
transformation
\begin{eqnarray}
\Gamma[\Phi] ~&=&~ W[{\cal J}] ~-~\int d^4 x~{\cal J} \cdot \Phi~ \nonumber \\
\Phi ~&=&~ {\delta W[{\cal J}] \over \delta {\cal J}} ~,
\end{eqnarray}
where, we have collectively labeled all background fields as $\Phi$ and the sources as
${\cal J}$, and $W[{\cal J}]$, we recall, is the
generating functional of connected Green's functions. 

The Effective Action can be derived iteratively from the integro-differential equation
\begin{eqnarray}
\exp({i\Gamma[\Phi]})~&=&~\int D\phi~ \exp\left({iS[\phi]+i\int d^4x~(\phi-\Phi)~\frac{\delta \Gamma}
{\delta\Phi}}\right)\label{iter}
\end{eqnarray}
where we have used the equation of motion for the effective action
\begin{eqnarray}
 \frac{\delta\Gamma[\Phi]}{\delta \Phi}=-{\cal J}(x,\Phi]
\end{eqnarray}
The r.h.s. of this equation is a function of configuration space variable $x$ but at the same 
time a functional of field $\Phi$.
The task is to
compute $\Gamma[\Phi]$ to ${\cal O}(\hbar)$ with a view to eventually obtaining
the one loop effective potential defined by the relation
\begin{eqnarray}
V_{eff}(\phi_0) ~\equiv ~-~ \Gamma(\Phi)|_{\Phi=\phi_0} ~\left(\int d^4x \right)^{-1} ~ ,
\label{veff}
\end{eqnarray}
where, $\phi_0$ are spacetime independent. Observe that $V_{eff}(\phi_0)$ is
the generating functional for 1PI graphs with vanishing external
momenta. Even though the scalar potential is classically scale invariant,
a mass scale is generated through renormalization in the quantum theory,
which breaks this scale invariance. The effective potential may thus have
a minimum away from the origin in $\rho$-space, defined in terms of the
renormalization mass scale, which, in turn, relates to values of the dimensionless
physical parameters of the theory (dimensional {\it transmutation} \cite{cw}).

Instead of evaluation of the functional integral over the $\Theta$ and ${\bf A}_{\cal
P}$ fields, we make a change of basis to $\Theta$ and ${\bf Y}$ via
(\ref{ugau}) and make use of
the action (\ref{abhm}) which is independent of $\Theta$. The latter appears
only in the constraint which now becomes a statement of non-transversality in
spacetime of the ${\bf Y}$ field. $\Theta$ can be simply integrated out,
leaving behind a field-independent normalization which we set to unity.
The integration over $\rho$
involves a saddle-point approximation around a field $\rho_c$ which may be
called a `quantum' field, since it is the solution of the classical
$\rho$-equation of motion augmented by ${\cal O}(\hbar)$ corrections. With
no gauge ambiguities anywhere, there is no question of gauge fixing; functional
integration over the physical vector potential ${\bf Y}$ can be performed
straightforwardly.

Following ref. \cite{jacki}, the one loop effective action is given
schematically by
\begin{eqnarray}
\Gamma^{(1)}~[\rho_c] ~=~ S[\rho_c,0,0] ~-~ i\hbar~Z^{(1)}[\rho_c] ~,
\label{olga}
\end{eqnarray}
where,
\begin{eqnarray}
Z^{(1)}[\rho_c]\equiv  \int {\cal D} \rho {\cal D}{\bf Y} ~\exp {i \over 2\hbar}
\left[~\int d^4xd^4y~\rho(x) {\cal M}_{\rho\rho}(x,y)~\rho(y) + Y^{\mu}(x) {\cal
M}_{Y_{\mu}Y_{\nu}}(x,y) Y^{\nu}(y)~ \right]~, \label{zone}
\end{eqnarray}
with, generically,
\begin{eqnarray}
{\cal M}_{AB}(x,y) ~ \equiv ~ \left({\delta^2 S[\Phi] \over \delta
\Phi_A(x)~\delta \Phi_B(y)}\right)_{\Phi=\rho_c,0,0}~. \label{emm}
\end{eqnarray}
Since our object of interest is the one loop effective potential, we restrict
ourselves to a saddle point $\rho_c$ which is spacetime independent.
The matrices ${\cal M}$ turn out to be diagonal in field space for the purpose of a one loop
computation, with entries
\begin{eqnarray}
{\cal M}_{\rho \rho}~ &=&~ - \left(~\Box ~+~ {\lambda \over 2} \rho_c^2
  ~\right) ~\delta^{(4)}(x-y) \nonumber \\
{\cal M}_{Y_{\mu} Y_{\nu}}~ &=&~\left[ \eta_{\mu \nu}~\left(
  ~\Box~+~e^2~\rho_c^2\right) - \partial_{\mu} \partial_{\nu} \right]\delta^{(4)}(x-y) ~.
\end{eqnarray}
One obtains easily
\begin{eqnarray}
Z^{(1)}[\rho_c]~&=&~ \left(~Det \left[ ~{\cal M}_{\rho \rho} {\cal M}_{YY}
    ~\right] \right)^{-1/2}~ ,\label{zloop}
\end{eqnarray}

The functional determinants are evaluated in momentum space following
\cite{jacki}, and one obtains for the one loop
effective potential, using eq. (\ref{veff}), the expression
\begin{eqnarray}
V_{eff}(\rho_c) &=& \frac{1}{4!} \lambda \rho_c^4~+~i\hbar
\int d^4k~\log \left [ \left(-k^2 + e^2 \rho_c^2 \right)^{3/2} \left
(-k^2 + \lambda \rho_c^2 \right)^{1/2} \right] \nonumber \\
&+& \frac12 B \rho_c^2 ~+~ \frac{1}{4!} C \rho_c^4 ~,
\end{eqnarray}
where $B$ and $C$ are respectively the mass and coupling constant
counterterms. The momentum integral is performed with a
Lorentz-invariant cut-off $k^2 = \Lambda^2$, yielding
\begin{eqnarray}
V_{eff}(\rho_c) &=& \frac{1}{4!}\rho_c^4~+~\frac12 B \rho_c^2 ~+~ \frac{1}{4!}
C \rho_c^4 \nonumber \\ 
&~+~& {\hbar \rho_c^2 \Lambda^2 \over 32 \pi^2} (\frac12 \lambda + 3
e^2) \nonumber \\
&+& {\hbar \rho_c^4 \over 64\pi^2} \left[\frac14 \lambda^2
  \left(\log {\lambda \rho_c^2 \over 2 \Lambda^2} - \frac12 \right)
+ 3 e^4 \left(\log {e^2 \rho_c^2 \over \Lambda^2} - \frac12 \right)
\right]\label{effpot}
\end{eqnarray}

We remark here that in these manipulations, a $\exp (-\log \rho^2)$ term is
generated in the one loop partition function, which cancels {\it exactly}
against an identical term $Det \rho^2$ arising in the formal measure as
discussed after eq. (\ref{vpam}). This is precisely the point that was made
earlier: the interpretation of that extra local factor in the formal
functional measure as some sort of conformal mode in a conformally flat
background is subject to some modification at the one loop level, since that factor is
{\it eliminated} by a one loop contribution to the partition function. This
has been anticipated in ref. \cite{rys} where an attempt has been made to give
an alternate interpretation in terms of a `gauge-dependent gravity'. Perhaps one
can use compensator fields to account for this loss of scale invariance due
to renormalization effects, in order to resurrect the novel interpretation
proposed in \cite{nie}, \cite{fadd}. 
  
The mass and coupling constant renormalizations $B$ and $C$ are fixed through
the renormalization conditions
\begin{eqnarray}
{d^2 V \over d\rho_c^2}\Bigg|_{\rho_c=M} &=& 0 \\
{d^4 V \over d\rho_c^4}\Bigg|_{\rho_c=M} &=& \lambda
\label{renc}
\end{eqnarray}
leading to the renormalized one loop effective potential

\begin{eqnarray}
V_{eff}(\rho_c) &=& {\lambda \over 4!} \rho_c^4 + \rho_c^2 M^2 \left[-{\lambda
    \over 4} + {9 \over 32\pi^2} (3e^4 + \frac12 \lambda^2) \right] \nonumber  \\ 
&+& \left({3e^4 \over 64 \pi^2} + {\lambda^2 \over 256 \pi^2} \right)\hbar \rho_c^4
\left[\log {\rho_c^2 \over M^2} - {25\over 6} \right]~.
\label{rveff}
\end{eqnarray}
In the usual approach, with or without the presence of $\theta$ (The latter case is usually regarded as ``unitarity gauge'')as a 
dynamical variable in the action of scalar-QED the expression of
one-loop effetive potential does not agree \cite{Dolan_Jackiw} with each other. This is not surprising to have different 
results for 
the two cases. If we carefully look into these two theories then we can find that in the latter, there
 the vertex  $\rho^2~\partial_{\mu}\theta A^{\mu}$
is absent. Therefore, it is obvious that the one-loop calculation starting from these
 two actions lead us to different results. However,
in gauge-free approach the calculation from two actions (\ref{abh2}) and (\ref{abhm}) yields same result.
 
The potential has an extremum at $\rho_c = \langle \rho \rangle(M)$ leading
eventually to the ratio of the squared masses of the Higgs boson to the photon
\begin{eqnarray}
{m_H^2 \over m_A^2} ~=~ \frac{1}{e^2} ~\left[\frac13 \lambda -\left({3e^4
      \over 8 \pi^2} + {\lambda^2 \over 32\pi^2} \right) \log {\langle \rho
      \rangle^2 \over M^2 } -{e^4 \over \pi^2} - {\lambda^2 \over 12\pi^2} \right]
\label{masr}
\end{eqnarray}
The derivation of the mass ratio of the Higgs mass to the photon seemingly
went through without any gauge fixing, since all fields being functionally
integrated over are {\it physical} fields without any gauge ambiguity. The
result (\ref{masr}) is thus a `physical' result in this toy model where the
photon acquires a mass. Notice that unlike in the original Coleman-Weinberg
paper, we did not make an approximation of choosing $\lambda \sim e^4$, to
drop terms of $O(\lambda^2)$. Thus, even though our result agrees with the
earlier papers qualitatively, there are significant quantitative
differences. However, the point in this section is not so much the result of
the computation of the mass ratio, but the observation that the effect
is physical and not a gauge artifact.

\section{DeWitt-Vilkovisky Effective Action}

We give a brief outline of the Vilkovisky's unique effective action here. It was pointed out by DeWitt that we must evaluate 
the effective action more carefully by treating the space of field configurations as a manifold. The problem with conventional 
formalism of effective action lies in the fact that the 
eqn. (\ref{iter}) 
\begin{eqnarray}
\exp{({i\Gamma[\Phi]})}~ = ~\int D\phi~ \exp{\left({iS[\phi]+i\int d^4x~(\phi-\Phi)~\frac{\delta \Gamma}{\delta\Phi}}\right)}\nonumber
\end{eqnarray}
does not have a correct geometrical interpretation because the
difference of two points in the configuration space ${\cal M}$, namely 
$\phi-\Phi$, in general, is not a vector on that space. This spoils the covariance of the expression 
$(\phi-\Phi)\frac{\delta \Gamma}{\delta \Phi}$ under field
reparameterization s. Thus the effective action fails to be a scalar
function on the configuration space ${\cal M}$. To get rid of this problem Vilkovisky
proposed that $\phi-\Phi$ should be replaced by a two-point function $\sigma^i(\Phi,\phi)$ [we adopt DeWitt's condensed notation here], 
which is a vector tangent to the curve joining 
the points $\Phi$ and $\phi$. It is a vector with respect to the point $\Phi$ and a scalar w.r.t. the point $\phi$. The properties of 
this bi-vector is discussed in detail in \cite{dewittdyn}. With this definition of $\sigma^i(\Phi,\phi)$ the 
effective action now been derived from 
\begin{eqnarray}
 \exp{(i\Gamma[\Phi])}~&=&~\int D\phi\, d\mu[\phi]~\exp{\left(iS[\phi]+i\int d^4x~\sigma^i(\Phi,\phi)~\frac{\delta \Gamma}{\delta\Phi}\right)}
\label{vdea}
\end{eqnarray}
Now the r.h.s. of eqn. (\ref{vdea}) becomes a scalar function of $\phi$ and the functional integral is independent of 
reparameterization
of $\phi$. We can expand $\sigma^i(\Phi,\phi)$ in powers of $\Phi-\phi$ with co-efficients evaluated at $\Phi$.

\begin{equation}
 \sigma^i(\Phi,\phi)=(\Phi-\phi)^i-{1\over2} \varGamma^i_{mn}(\Phi)(\phi-\Phi)^m(\phi-\Phi)^n+\cdots
\end{equation}

In one-loop approximations one gets

\begin{equation}
 \Gamma_{DV}^{(1)}[\Phi]=S(\Phi)+{\hbar \over i}~\ln\mu[\phi]+{\hbar \over 2i}Tr\ln[\nabla_m\nabla_nS(\Phi)]+O(\hbar^2) 
\end{equation}

 where $\,\nabla_{m}\,$ is the covariant derivative associated  with
 connection $\,\varGamma^i_{mn}$. The connection is Christoffel and completely 
described by the metric on the manifold ${\cal M}$.

 \begin{equation}
                  \,\varGamma^i_{mn}={1 \over 2}G^{ik}(G_{mk,n}+G_{nk,m}-G_{mn,k})\
                 \end{equation}

and
 \begin{equation}
 \frac{\delta^2S}{\delta\phi^m\delta\phi^n}
 ~\to~
 \nabla_m\nabla_nS=\frac{\delta^2S}{\delta\phi^m\delta\phi^n}-\varGamma^i_{mn}(\phi)
 \frac{\delta S}{\delta\phi^i}\,.
 \end{equation}

For gauge theories we must evaluate the effective action in physical configuration space i.e. the space of all the gauge fields modulo
the possible gauge transformations. First consider the infinitesimal gauge transformation,
 \begin{equation}
 \delta\phi^{i} ~=~ K^{i}_{\alpha}[\phi]\epsilon^{\alpha}\,,
 \end{equation}
 with $K^{i}_{\alpha}[\phi]$ being the generators of gauge transformation
 and $\epsilon^{\alpha}$ the infinitesimal gauge-group parameters.

Let $G_{ij}$ be the metric in the naive field space;

\begin{equation}
 ds^2=G_{ij}\delta\phi^i\delta\phi^j
\end{equation}

The metric of the physical field space is given by,

\begin{equation}
 ds^2_{P}=\gamma_{ij}\delta\phi^i\delta\phi^j=G_{ij}\delta\phi_P^i\delta_P^j
\end{equation}

where the physical field is defined as 
\begin{equation}\delta\phi_P^i=\Pi^i_j\delta\phi^j\end{equation} 

The projector $\Pi^i_j$ projects the vectors of naive 
field space onto a subspace of vectors which are perpendicular to the space of tangent vectors to the orbits generated by 
$K^i_{\alpha}$.

\begin{equation}
 \Pi^i_j=\delta^i_j-K^i_{\alpha}N^{\alpha\beta}K^l_{\beta}G_{lj}
\end{equation}

with 

\begin{equation}
 N^{\alpha\beta}=\gamma^{\xi\alpha}\gamma^{\chi\beta}K^i_{\xi}K^j_{\chi}G_{ij}
\end{equation}

The modification of the connection due to gauge field gives

\begin{equation}
 \Gamma^i_{jk}=\varGamma^i_{jk}~+~T^i_{jk}
\end{equation}
where we have ignored a piece proportional to $K^i_{\alpha}$ since it will annihilate the action and will not contribute to the 
one-loop order. The gauge-part of connection reads,

\begin{equation}
 T^i_{jk}=K^{\alpha}_{(i}K^{\beta}_{j)}K^l_{\alpha}K^m_{\beta;l}~-~K^{\alpha}_iK^m_{\alpha;j}~-~K^{\alpha}_jK^m_{\alpha;j}
\end{equation}

\subsection{Scalar QED in Gauge-free DV approach}

The detailed calculation of one-loop effective potential for scalar
QED is given in \cite{Kun}. We do not repeat it here but merely
restate essential results of that work. The action for the scalar QED is 
\begin{eqnarray}
S[\rho,\theta, A] = \int d^4x \left[ \frac12 (\partial
\rho)^2 + \frac12 e^2 \rho^2 ( A - e^{-1}\partial \theta)^2
~- \frac12 (\partial A)^2 -  {\lambda \over 4!}\rho^4 \right], 
\end{eqnarray}
The metrics in field space are given by
\begin{eqnarray}
G_{\rho(x)\rho(y)}~&=&~\delta^4(x-y)\\ 
G_{\theta(x)\theta(y)}~&=&~\rho^2\delta^4(x-y)\\
G_{A_{\mu}(x)A_{\nu}(y)}~&=&~-\eta_{\mu\nu}\delta^4(x-y)
\end{eqnarray}

For scalar QED the correct measure which is invariant under general co-ordinate transformations in ${\cal M}$ will contain
the determinant of the metric. Thus here,
\begin{equation}
d\mu[\phi]=\sqrt{det G}=Det|\rho(x)\delta^4(x-y)|\nonumber 
\end{equation}
The only contribution to the one-loop effective potential from the Christoffel symbol $\varGamma^i_{jk}$ is
\begin{eqnarray}
 \varGamma^{\rho(z)}_{\theta(x)(\theta(y)}~&=&~-\frac{\lambda\rho_c^4}{6}\delta^4(x-y)\\
\end{eqnarray}

To get the contribution from gauge part of the connection $T^i_{jk}$ we identify the generators of gauge transformations
\begin{eqnarray}
K^{\rho(x)}_y~&=&~0\\
K^{\theta(x)}_y~&=&~e\delta^4(x-y)\\
K^{A_{\mu}(x)}_y~&=&~-\partial_{\mu}\delta^4(x-y)
\end{eqnarray}
with the partial derivative with respect to the first argument of the $\delta$-function. We write down the non-trivial
 $\varGamma^i_{jk}$s. The calculation details can be found in \cite{Kun}.
\begin{eqnarray}
T^{\rho(z)}_{\theta(x)\theta(y)}&=&e^2\rho^3_c[\delta^4(x-y)N^{zx}~+~\delta^4(z-x)N^{zy}-e^2\rho^2_cN^{xy}N^{xz}]\\
T^{\rho(z)}_{A_{\mu}(x)A_{\nu}(y)}&=&-e^4\rho_c^5\partial^{\mu}N^{yx}\partial^{\nu}N^{zx}\\
T^{\rho(z)}_{A_{\mu}(x)\theta(y)}&=&e\rho_c[\partial^{\mu}N^{xz}\delta^4(x-y)-e^2\rho_c^2\partial^{\mu}N^{xy}N^{xz}]
\end{eqnarray}
The one-loop effective potential calculated in this formalism turns out to be independent of the gauge parameter. In fact
it is equal to the one calculated by Jackiw with $\alpha=-1$ \cite{Kun}.
\begin{eqnarray}
V_{eff}(\rho_c) &=& {\lambda \over 4!} \rho_c^4 + {\hbar \over 64\pi^2} \left(3e^4 + \frac{5}{18} \lambda^2+ \frac23 \lambda e^2 \right)\rho_c^4\left[\log {\rho_c^2 \over M^2} - {25\over 6} \right]~.
\label{rveff}
\end{eqnarray}

Now, we turn to the case of gauge-free scalar QED. In our gauge-free approach the action from which
we have calculated the effective potential is given by eqn. (\ref{abh2}), 
\begin{eqnarray}
S[\rho,\Theta,{\bf A}_{\cal P}] = \int d^4x \left[ \frac12 (\partial
\rho)^2 + \frac12 e^2 \rho^2 ({\bf A}_{\cal P} - e^{-1}\partial \Theta)^2
~- \frac12 (\partial A_{\cal P})^2 -  {\lambda \over 4!}\rho^4 \right], \nonumber
\end{eqnarray}

Now for reparameterization  invariance we apply DV technique to calculate the effective potential 
for this action. The metrics of the field space are
\begin{eqnarray}
G_{\rho(x)\rho(y)}~&=&~\delta^4(x-y)\\ 
G_{\Theta(x)\Theta(y)}~&=&~\rho^2\delta^4(x-y)\\
G_{A_{\mu}(x)A_{\nu}(y)}~&=&~-\eta_{\mu\nu}\delta^4(x-y).
\end{eqnarray}

However, the only non-trivial contribution to the one-loop effective potential will be from $\varGamma^{\rho}_{\Theta~\Theta}$ and
an additional contribution occurs from our gauge-free conventional calculation.
\begin{equation}
 {\cal M}_{\Theta(x)\Theta(y)}=-\rho_c^2\left[-k^2+\frac{\lambda\rho^2_c}{6}\right]
\end{equation}

Since again the theory doesn't posses any non-vanishing gauge generators $(K^{\Theta(x)}_y=0; K^{{\bf A}_P(x)}_y=0)$ we don't have any 
gauge part of the connection. The one-loop effective potential in this gauge-free framework becomes:

\begin{eqnarray}
 V_{eff}=&&\frac{\lambda \rho_c^4}{4!}~-~i\hbar \int \frac{d^4 k}{(2\pi)^4}\log{\rho_c}~+~{i\hbar \over 2}\int \frac{d^4 k}{(2\pi)^4}\log[{-\rho_c^2k^2}]\nonumber\\
~&+&~{i\hbar \over 2}\int \frac{d^4 k}{(2\pi)^4}\log (-k^2+\frac{\lambda\rho^2_c}{6}) \nonumber \\ 
&+& ~i\hbar
\int d^4k~\log \left [ \left(-k^2 + e^2 \rho_c^2 \right)^{4/2} \left
(-k^2 + \lambda \rho_c^2 \right)^{1/2}\right] \nonumber\\~&+&~{\frac12}i\hbar Tr\int d^4k \log\left[\frac{k^2}{k^2-e^2\rho_c^2}\right]
\end{eqnarray}

The first integral comes from the integration measure which exactly cancels a divergent part coming from the inverse propagator 
of $\Theta$(the second integral). The last term is the contribution from the transversality constraint on ${\bf A_P}$, as already
included in eqn. (\ref{vpam}). This result clearly differs from the earlier result calculated by Kunstatter in \cite{Kun}. 
It also doesn't agree with the result obtained by gauge-free approach (\ref{effpot}) due to the third integral in the r. h. s. of 
$V_{eff}$. This is due to the fact that we have ignored the reparameterization 
invariance of the gauge-theories which was not captured in the gauge-free non geometric approach. The renormalized one-loop effective potential 
becomes

\begin{eqnarray}
V_{eff}(\rho_c) &=& {\lambda \over 4!} \rho_c^4 + {1 \over 64 \pi^2}\left(3e^4  + {5\lambda^2 \over 18} \right)\hbar \rho_c^4
\left[\log {\rho_c^2 \over M^2} - {25\over 6} \right]~.
\label{epvdgf}
\end{eqnarray}

This is the unique gauge-free Coleman-Weinberg potential for scalar QED and surprisingly this coincides with the result of 
Coleman-Weinberg's original paper! This is just a coincidence because this calculation doesn't involve any gauge fixing in oppose
to the case of Coleman-Weinberg's paper. So this is indeed a unique result which is free of any background or fluctuation gauge
ambiguities and also invariant under field reparameterization s.

\section{Generalization}
\subsection{Kalb-Ramond two form potential}

The Kalb-Ramond two form potential ${\bf B}$ has a field strength ${\bf H} =
d{\bf B}$ which is clearly invariant under the gauge transformation ${\bf B}
\rightarrow {\bf B} + d\Lambda$ for any one form field $\Lambda$. Construct
now the projected two form field ${\bf B}^T \equiv {\cal P} \otimes {\cal P}
{\bf B}$. Since $ {\cal P} d f=0~ \forall f$, under the
gauge transformation of ${\bf B}$, ${\bf B}^T \rightarrow {\bf B}^T + {\cal
P}\otimes {\cal P} d \Lambda = {\bf B}^T$. Further, in a coordinate system,
\begin{eqnarray}
\partial _{\mu} ~B^{T \mu \nu}~=~ 0 \label{trkr}
\end{eqnarray}
implying that it is indeed transverse. Finally, it is clear that ${\bf H} = d
{\bf B} = d{\bf B}^T$, which means that ${\bf B}^T$ is indeed the physical part of
the two form potential. 

As in the case of gauge free electrodynamics, one can formulate
the theory of Kalb-Ramond fields purely in terms of a {\it physical}
antisymmetric tensor potential $B_{\mu \nu}$ defined by the action 
\begin{eqnarray}
S_{KR}~=~\int d^4x \left( -\frac12 B_{{\cal P} \nu \rho} ~\Box B_{\cal P}^{\nu
    \rho} + J_{\nu \rho} ~B_{\cal P}^{\nu \rho} \right)  ~,
\label{kract}
\end{eqnarray}
where, $\partial^{\mu} B_{{\cal P} \mu \nu} =0 = \partial^{\mu} J_{\mu
\nu}$. 

We once again ask how unique the potential $B_{\cal P \mu \nu}$ is. Observe that both
the field equation and the divergenceless condition remain invariant under a gauge
transformation $B_{\cal P \mu \nu} \rightarrow (B_{\cal P \mu \nu})^{\Lambda}
= B_{\cal P \mu \nu} + 2\partial_{[\mu} \Lambda_{\nu]}$ where $\Lambda_{\mu}$
satisfies the equation $\Box \Lambda_{\mu} - \partial_{\mu} \partial \cdot
\Lambda = 0$. In contrast to the case of the graviton field, it is obvious
that this equation has an infinity of {\it gauge equivalent} solutions, the
equivalence being under $\Lambda_{\mu} \rightarrow \Lambda_{\mu}
+ \partial_{\mu} \omega$ for an arbitrary function $\omega$. Restricting
$\Lambda_{\mu}(\infty) = 0$ is not enough to make it vanish everywhere. We
need to additionally restrict $\partial \cdot \Lambda =0$ everywhere with the
requirement that $\omega(\infty) = const$. This additional restriction appears
necessary in this preliminary investigation to make the two form potential
unique. 

The reason why an identical procedure as for the photon or graviton field
does not suffice to yield a gauge-free formulation of antisymmetric tensor
potentials is because of the aspect of {\it reducibility} of these potentials: the
vectorial gauge parameter of the two form potential itself has a gauge
invariance. Perhaps our approach will need to be somewhat modified to produce
a gauge-free theory of potentials that have a reducible gauge invariance. 

\section{Conclusion}

Generalization of the foregoing approach to Yang Mills theories,
as has already been mentioned, has been achieved in the context of the electroweak
theory where Higgs scalars are assumed to be present \cite{nie, fadd}. In
these papers, a residual $U(1)$ gauge theory corresponding to the Maxwell
theory has been obtained. We have already been succeeded to wipe out the residual freedom from the theory and to rewrite it in terms of completely gauge-inert variables \cite{bm2}. However, a comprehensive study of all quantum 
properties of such a formulation is
under way and will be reported elsewhere.

For pure Yang Mills theories, the construction of a gauge-free alternative has
not yet been attempted, even though lattice gauge theories represent an
explicitly gauge invariant formulation. A local, gauge-free formulation of
Yang Mills theories is not obviously in contradiction with extant ideas about
colour confinement of quarks and gluons. This gives us the opportunity to
attempt a construction of a physical {\it non-Abelian}
one form in terms of the usual Yang-Mills gauge one form ${\bf A}$ (which takes
values in the Lie algebra of the gauge group ${\cal G}$). 

Defining the holonomy along
the curve $C$ from $y$ to $x$ as $h_{C[y,x]}[{\bf A}] \equiv {\bf P} \exp \int_{C(y,x)} {\bf
A}$, with ${\bf P}$ denoting path ordering, we note that under local gauge
transformations of the gauge potential $[{\bf A}(x)]^{\Omega(x)} =
\Omega(x)^{-1}[{\bf A}(x) + d] \Omega(x)$, where $\Omega \in {\cal G}$
the holonomy variables transform as 
\begin{eqnarray}
 h_{C(y,x)}[{\bf A}^{\Omega}]  = \Omega^{-1}(y) ~h_{C(y,x)}[{\bf A}]~ \Omega(x)
 ~. \label{holo}
\end{eqnarray}
If we choose the point $y \rightarrow \infty$ and require $\Omega(\infty) =
{\cal I}$, eqn. (\ref{holo}) now takes the form 
\begin{eqnarray}
h_{C(\infty, x)}[{\bf A}^{\Omega}] = h_{C(\infty, x)}[{\bf A}]~\Omega(x)
~.\label{grv}
\end{eqnarray}

We now formally define a {\it local} one form potential ${\cal A}(x)$ as
\begin{eqnarray}
{\cal A}(x) \equiv \int {\cal D}C~{\bar {\bf A}}_{C(\infty,x)} 
\end{eqnarray}
where, 
\begin{eqnarray}
{\bar {\bf A}}_{C(\infty,x)} \equiv h_{C(\infty,x)}[{\bf A}] \left( {\bf A} +
  d \right)  (h_{C(\infty,x)}[{\bf A}])^{-1}. ~\label{abar}
\end{eqnarray}
The path integral symbol at this point is formal, and is meant to stand for
some sort of averaging over all paths originating at asymptopia and extending
upto the field point $x$. It is then easy to see that, under gauge
transformations of ${\bf A}$ and
using eqn. (\ref{grv}), 
\begin{eqnarray}
{\cal A}^{\Omega}(x) ~=~ {\cal A}(x) ~. \label{gf}
\end{eqnarray}
What we have not been determined yet is what constraint replaces the
divergencefree condition (\ref{trans}) for the Yang Mills one form ${\cal A}$, so that the
physics of these local gauge-free one forms can be explored
more thoroughly without gauge encumbrances. One also envisages application of these ideas to general
relativity formulated as a gauge theory of Lorentz (or Poincar\'e)
connection. We hope to discuss these and consequent issues elsewhere.  
\vglue .5cm

\noindent {\bf Acknowledgment :} We thank R. Basu, A. Chatterjee, A. Ghosh, B.
Sathiapalan and S. SenGupta for useful discussions.

\end{document}